\documentclass[12pt]{article}
\usepackage{siunitx,color,amsmath,url,graphicx,authblk}

\begin{document}

\title{Review of the AC loss computation for HTS using $H$ formulation}

\author[1]{Boyang Shen}
\author[2]{Francesco Grilli}
\author[1]{Tim Coombs}
\affil[1]{\small Electrical Engineering Division, Department of Engineering, University of Cambridge, Cambridge CB3 0FA, United Kingdom}
\affil[2]{\small Institute for Technical Physics, Karlsruhe Institute of Technology, 76131 Karlsruhe, Germany}
\date{}
\maketitle

\begin{abstract}
This article presents a review of the finite-element method (FEM) model based on the $H$ formulation of Maxwell's equations used to calculate AC losses in high-temperature superconductor (HTS) tapes, cables and windings for different applications. This model, which uses the components of the magnetic field as state variables, has been gaining a great popularity and has been in use in tens of research groups around the world.
This contribution first reviews the equations on which the model is based and their implementation in FEMfinite-element programs for different cases, such as 2D longitudinal and axis-symmetric geometries, and 3D geometries. Modeling strategies to tackle large number of HTS tapes, such as multi-scale and homogenization methods, are also introduced.
Then, the second part of the article reviews the applications for which the $H$ formulation has been used to calculate AC losses, ranging from individual tapes to complex cables and large magnet windings.
Afterwards, a section is dedicated to the discussion of the $H$ formulation in terms of computational efficiency and easiness of implementation. Its pros and cons are listed. Finally, the last section draws the main conclusions.
\end{abstract}

\section{Introduction}

High-temperature superconductors (HTS) are recognized as the solution for future superconducting applications, because of their high current and power density, good in-field behavior, and mechanical strength~\cite{Larbalestier:Nat01,Hassenzahl:PIEEE04}. 
In recent years, the  performance of HTS has been enhancing, while their price has been gradually decreasing, and several types of conductors have been proposed for different applications~\cite{Parizh:SST17,Uglietti:SST19}.
All these advantages make HTS a technically viable solution for future superconducting power applications, such as high-field magnets~\cite{Hahn:Nature19}, motors and generators~\cite{Snitchler:TAS11,Huang:TAS16}, transformers~\cite{Schwenerly:TAS99,Iwakuma:TAS01},  power transmission cables~\cite{Maguire:TAS07},  fault current limiters~\cite{Noe:SST07}.

AC losses, however, represent an important  limiting factor for the commercialization of HTS, especially when solutions based on conventional materials are available. One has to remember that superconductors operate at very low temperatures, and that a power dissipation that would not constitute a concern at room temperature can represent a very serious burden when refrigeration costs are taken into account. For example, the Carnot specific power for a temperature of \SI{77}{\kelvin} (boiling temperature of liquid nitrogen at atmospheric pressure) is about 2.9, but realistic estimates must take into account the efficiency of the cooling systems, which is in the range of 10-\SI{20}{\percent}~\cite{Wang:Cryocoolers08}.
Therefore, scientists and engineers need to estimate the amount of AC losses and take the corresponding actions. Different types of models can be used for this purpose.

\begin{figure}[t!]
	\centering
	\includegraphics[width=8 cm] {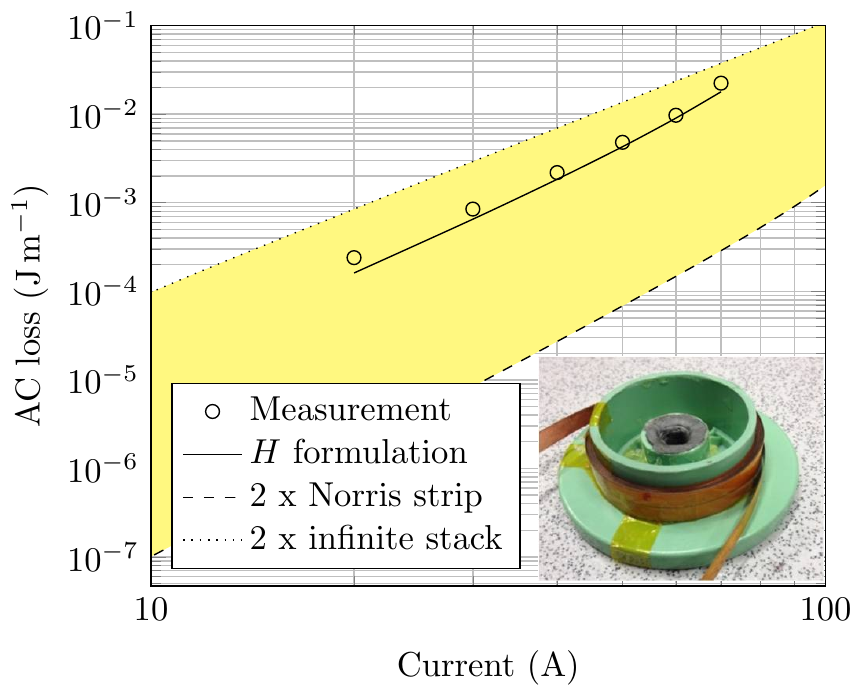}
	\caption{\label{fig:Boyang_plot}Transport AC losses of an HTS double pancake coil ($2 \times 18$ turns). Experimental results are compared to those calculated with the $H$ formulation and with analytical models for a single tape~\cite{Norris:JPDAP70} and an infinite stack of tapes~\cite{Muller:PhysC97a,Muller:PhysC99}. The shaded area emphasizes the several orders of magnitude of difference for the losses calculated with the two analytical models. Data taken from~\cite{Shen:SST18}.}
	\end{figure}

In the past decades, several analytical models have been developed for calculating  AC losses in superconductors by means of simple mathematical formulas. Notable examples are the expressions for the transport losses of superconducting tapes with elliptical or infinitely thin cross section~\cite{Norris:JPDAP70} and the magnetization losses of superconducting tapes with infinitely thin cross section~\cite{Halse:JPDAP70,Brandt:PRB93,Zeldov:PRB94}. In the case of infinitely thin tapes, expressions for AC losses have been derived also for infinite stacks or arrays with transport current or subjected to an external magnetic field~\cite{Mawatari:PRB96,Muller:PhysC97a,Muller:PhysC99,Mawatari:PRB08}, and for individual tapes with substrate of infinite magnetic permeability. A comprehensive review of these and other expressions is given in~\cite{Mikitik:TAS13}.
Typical limitations of these analytical models are the restriction to simple geometries and the use of the critical state model as constitutive law of the superconductors. Numerical models, such as the FEM model based on the $H$ formulation of Maxwell's equations that is the topic of this review, allow overcoming these limits and investigating the electrodynamic response (and in particular the AC losses) of realistic applications.  
As an example, figure~\ref{fig:Boyang_plot} shows the transport AC losses of an HTS double pancake coil. Analytical models can calculate the losses of a single turn~\cite{Norris:JPDAP70} or of an infinite number of turns~\cite{Muller:PhysC97a,Muller:PhysC99}: these estimations can only provide lower and upper limits for the AC losses and are not accurate for a real coil made of a finite number of turns. Numerical models such as the $H$ formulation can simulate the realistic geometry, thus providing an estimation of the AC losses much closer to the measured data.

The idea of simulating the electromagnetic behavior of superconductors in time-dependent problems by using finite elements with the magnetic field as state variable was first proposed in 2003 by Kajikawa et al.~\cite{Kajikawa:TAS03} and Pecher et al.~\cite{Pecher:EUCAS2003}. In both cases, the formulation was implemented in home-made codes.
A few years later, Hong~et al.~\cite{Hong:SST06} and Brambilla~et al.~\cite{Brambilla:SST07} independently implemented the $H$ formulation in the commercial FEM software COMSOL Multiphysics~\cite{Company:Comsol}. This model became quickly very popular and it is now the de facto standard in the applied superconductivity community. A search on the Web of Science based on the citations of~\cite{Hong:SST06,Brambilla:SST07} revealed that, at the time of writing, the $H$ formulation has been used by at least 45 research groups worldwide. 

This review article is organized as follows.
First, in section~\ref{sec:Hform}, the mathematical implementation of the $H$ formulation is discussed, first for its basic form, then for its extension and adaption for solving problems of increasing complexity. Then, in section~\ref{sec:applications}, the application of the $H$ formulation to cases of practical interest is reviewed. Further, section~\ref{sec:discussion} is dedicated to discussing the reasons of the popularity of this formulation, its implementation in different programming environment, its advantages and drawbacks. Finally, the main conclusions of this review article are summarized in section~\ref{sec:conclusion}.

This review is specifically dedicated to the use of the $H$ formulation for investigating the AC losses of HTS in a variety of geometric arrangements and working conditions at constant temperature. The coupling with thermal models is therefore not considered here.

\section{$H$ formulation}\label{sec:Hform}

\subsection{Basic equations}\label{sec:basic_equations}
The model solves Faraday's equation in a finite-element environment, using the magnetic field components as state variables
\begin{equation}\label{eq:faraday}
\nabla \times \bf{E}=-\frac{\partial \bf{B}}{\partial{t}}
\end{equation}
where $\bf{B}=\mu \bf{H}$, with $\mu=\mu_{\rm r} \mu_0$.\footnote{The first part of this section is mostly taken from~\cite{Zermeno:SST13}.} The lower critical field below which type-II superconductors are in the Meissner state is usually very low (\SI{}{\milli\tesla} range), so for most practical cases of power applications one can assume $\mu_{\rm r}$=1 for the superconductor material.

Since $\bf{E}=\rho \bf{J}$ and $\bf{J}=\nabla \times \bf{H}$, one can rewrite~(\ref{eq:faraday}) in terms of the magnetic field as
\begin{equation}\label{eq:faradayH}
\nabla \times (\rho \nabla \times \bf{H})=-\frac{\partial (\mu \bf{H})}{\partial{t}},
\end{equation}
where the magnetic field needs also to obey Gauss's law 
\begin{equation}\label{eq:gauss}
\nabla \cdot (\mu \bf{H})=0.
\end{equation}

The simultaneous solution of equations~(\ref{eq:faradayH}) and~(\ref{eq:gauss}) presents a problem, as it involves an over-constrained system in the general case. This issue was addressed by Bossavit and V\'erit\'e~\cite{Bossavit:TMAG83}, who wrote the equations in weak form and treated the problem as one of functional analysis. An elegant equivalent solution was presented by Kajikawa~et al.~\cite{Kajikawa:TAS03}: taking the divergence of equation~(\ref{eq:faradayH}) yields

\begin{equation}\label{eq:faradayH2}
\nabla \cdot \left [\nabla \times (\rho \nabla \times \bf{H})\right ]=\nabla \cdot \left ( -\frac{(\mu \bf{H})}{\partial t}\right ).
\end{equation}

The left-hand side of equation~(\ref{eq:faradayH2}) is identically zero and, after exchanging the order of time and spatial derivatives, it is easy to see that $\nabla \cdot \bf{B}=\nabla \cdot (\mu \bf{H})$ is constant in time. Consequently, if  $\nabla \cdot \bf{B}=0$ at a given time $t_0$, then $\nabla \cdot \bf{B}=0$ will hold at any other instant. So, if initial conditions are chosen such that 
\begin{equation}\label{eq:initcond}
\nabla \cdot (\mu \bf{H}) |_{t=t_0}=0
\end{equation}
then $\nabla \cdot (\mu \bf{H})=0$ will hold at all times.
It is important to remark that the divergence-free characteristic of the magnetic flux density $\bf{B}$ is enforced solely by analytical construction, independently of the type of elements used in the FEM implementation. However, this way to impose the divergence-free characteristic on $\bf{B}$ can be sensitive to errors if the time integration solver is not sufficiently robust, and makes some type of elements preferable to others. In particular, for the implementation of this model in the commercial software package COMSOL Multiphysics~\cite{Company:Comsol}, first-order edge elements are preferable. A study of the effect of different types (Lagrange or edge) and order (first or second) of elements is presented in~\cite{Ainslie:COMP11}.

\begin{figure}[t!]
	\centering
	\includegraphics[width=8 cm] {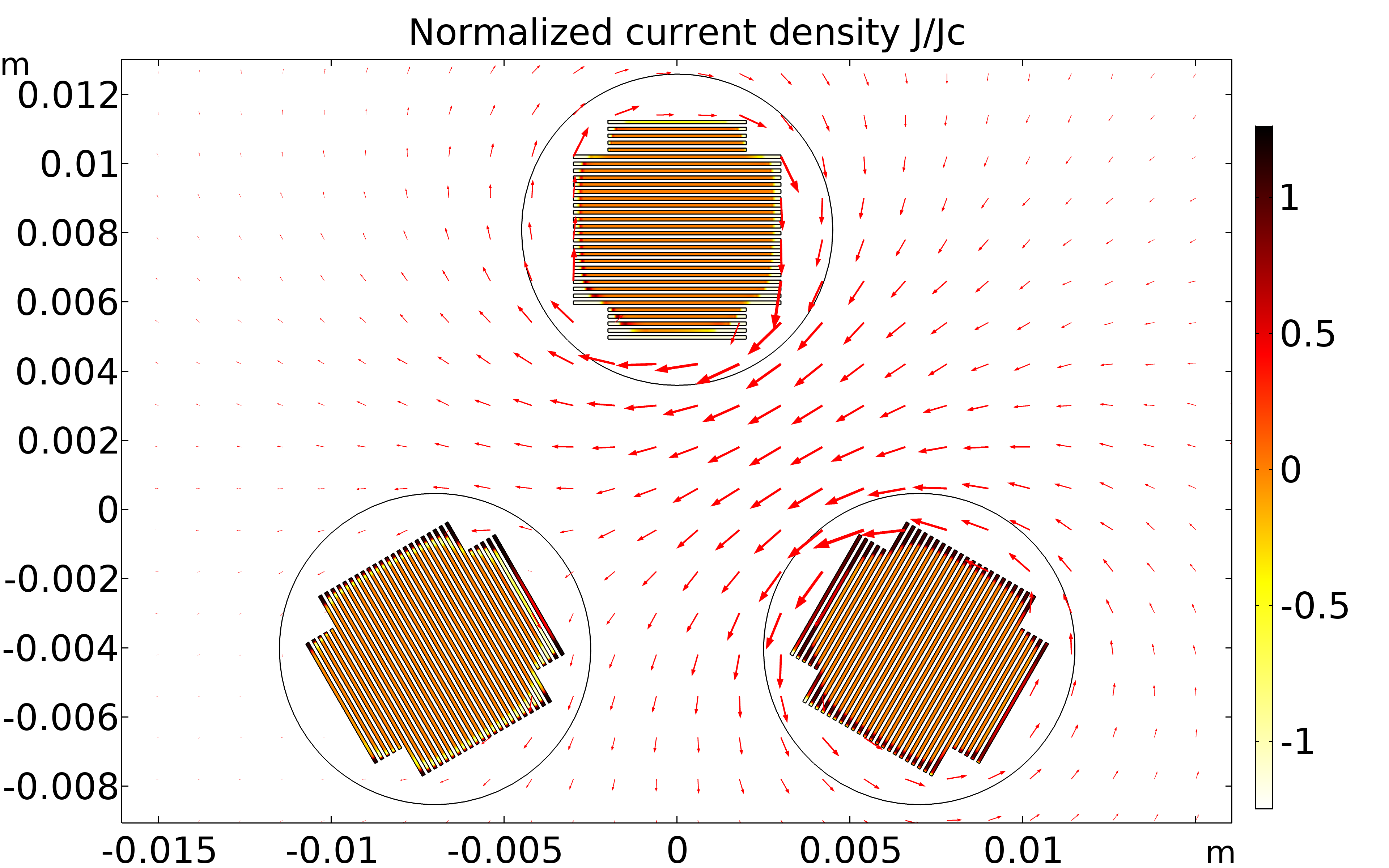}
	\caption{\label{fig:CroCo}{Normalized current density distribution of a three-phase Cross-Conductor (CroCo) cable~\cite{Shen:TAS19} at a given instant during the AC cycle.  Three \SI{120}{\degree}-shifted sinusoidal currents are imposed in each CroCo cable. Since such cables are made of non-transposed tapes, the current is let free to distribute among the 32 tapes composing each cable. If the tapes were fully transposed (such as in a Roebel cable), further current constraints would be needed in order to have all tapes carry identical currents. The arrows indicate the direction of the generated magnetic field, their magnitude being proportional to the field's magnitude.}}
	\end{figure}

A Dirichlet boundary condition allows modeling instances where magnetic field, transport current or a combination of both are considered. In the case of a transport current flowing in several conductors, a set of integral constraints allows fixing the net current in each conductor.
Therefore a net current $I_k(t)$ can be imposed in the $k^{\rm th}$ conductor by enforcing
\begin{equation}\label{eq:current_constraints}
I_k(t)=\int\limits_k \bf{J} \cdot {\rm{d}\bf{A}}_k,
\end{equation}
where $\bf{A}_k$ denotes any open surface that completely intersects the $k^{\rm th}$ conductor alone. Figure~\ref{fig:CroCo} shows an example of how current constraints can be used to control the current distribution in a three-phase Cross-Conductor (CroCo) cable.

The superconductor is modeled as a material with nonlinear electrical resistivity, usually in the form of a power-law
\begin{equation}\label{eq:PL}
\rho(J)=\frac{E_{\rm c}}{J_{\rm c}} \left ( \frac{J}{J_{\rm c}(\bf{B})}\right )^{n(\bf{B})-1},
\end{equation}
where $J$ is the magnitude of the current density, $E_{\rm c}$ is usually set equal to \SI{1e-4}{\volt\per\meter}, $J_{\rm c}$ is the superconductor's critical current density, and $n$ is the power-index describing the flux creep. Such a nonlinear resistivity mirrors the nonlinear voltage-current relationship observed in the characterization of HTS tapes~\cite{Rhyner:PhysC93}, where the critical current $I_{\rm c}$ is defined as the current at which a threshold voltage is reached.

In general, the critical current density $J_{\rm c}$ depends on the magnetic field amplitude and orientation~\cite{Grilli:SST10a,Robert:Mat19}, and this dependence can assume fairly complicated forms~\cite{Grilli:TAS14b}. 
The power index $n$ too depends on the magnetic field amplitude and orientation, although in most simulations its field dependence is disregarded, as it is expected to have a less important influence on the results than that of $J_{\rm c}$.
The dependence of $J_{\rm c}$ on the position inside the tape, for example as a result of the manufacturing process of HTS tapes~\cite{Grilli:TAS07a}, can also be easily included.
The space around the conductors is usually modeled as a material characterized by very high resistivity (e.g. \SI{1}{\ohm\meter}), so that no current flows there. In most cases, the tapes' or device's behavior is determined by the superconducting parts only, and the other materials composing the superconducting tapes are ignored. However, they can be easily included and assigned the proper values of electric resistivity and magnetic permeability.  In some cases, magnetic materials need to be modeled. This constitutes a challenge because their magnetic permeability is usually not constant and, in particular, it depends on the magnetic field. Since the magnetic field varies with time, this means that in equation~(\ref{eq:faradayH}) the term $\mu$ cannot be simply taken out of the time derivative.  As a consequence, the expression of Faraday's equation to be implemented in the model needs to be properly changed, as described in~\cite{Nguyen:SST10}. Details about the practical implementation in COMSOL Multiphysics, including hints to help convergence, are given in~\cite{Krueger:Thesis14}.

The model can also be modified to take into account the presence of resistive components in series with the superconductors~\cite{Zermeno:SST14b}. This is for example useful for simulating (in 2D) the current distribution in cables composed of several tapes, each characterized by a different termination resistance. In this case, an additional term $R_i I_i$ must be added to the electric field contribution of each superconductor, where $R_i$ and $I_i$ are the termination resistance and current of the $i^{\rm th}$ superconducting tape.
The final model therefore consists of equation~(\ref{eq:faradayH}), initial conditions fulfilling equation~(\ref{eq:gauss}) and a set of appropriate boundary conditions and/or constraints. One should note that the treatment above is invariant as it does not rely on the choice of any specific coordinate system or reduced dimensionality; therefore it holds for rectangular, cylindrical, spherical or any other coordinate system in one, two, or three dimensions.

The model can be used to compute AC losses of HTS tapes and devices in a variety of configurations and operating conditions. Typically, situations with an AC transport current, an AC magnetic field, or a combination of the two are simulated. A sinusoidal current and/or field excitation is imposed (with the appropriate set of current constraints and boundary conditions) and the AC losses are calculated by integrating the instantaneous local dissipation $\bf{J \cdot E}$ over the superconducting domain and averaging it over a cycle. For such calculation, the first quarter of the sinusoidal cycle must be avoided, because it is not representative of the AC regime due to the occurring transient. The cyclic AC losses can therefore be computed on the second half of the first cycle as
\begin{equation}\label{eq:losses}
Q=2 \int\limits_{1/(2f)}^{1/f} \int\limits_{\Omega} \bf{J \cdot E} ~{\rm d}\Omega~{\rm d}t
\end{equation}
where $f$ is the frequency of the AC source and $\Omega$ the superconducting domain. This expression can be adapted to calculate the losses for arbitrary time-dependent excitations.

\subsection{Equations in 2D for longitudinal and axisymmetric problems}\label{sec:2D}
A large number of applications can be modeled as a set of long straight conductors or as windings with cylindrical symmetry, for which a 2D description considering the transversal cross section of the superconductors is sufficient. In this case, the state variables are the two components of the magnetic field in the considered cross section; the current can only flow perpendicular to them. This 2D approximation automatically implies that the materials' properties are assumed to be constant along the conductors' length. Therefore, for example, the variation of $J_{\rm c}$ along a tape's length -- which is observed experimentally~\cite{Gomory:SST19} -- cannot be taken directly into account.
In the case of infinitely long conductors (like straight tapes or cables), it is sufficient to consider their transversal cross section ($xy$ plane). In this case, the magnetic field has only two components ${\mathbf H} = [H_x, H_y, 0]$. The current density only one component ${\mathbf J}=[0, 0, J_z]$, which, since $\bf{J}=\nabla \times \bf{H}$, can be expressed in 2D as 
\begin{equation}
J_z=\frac{\partial H_y}{\partial x}-\frac{\partial H_x}{\partial y}.
\end{equation}

In the case of conductors exhibiting cylindrical symmetry (like solenoids, pancake coils) the same concept applies. The only difference is that the equations need to be written in cylindrical coordinates.

In cylindrical coordinates ($r,\theta, z$), 
%
the three components of equation (\ref{eq:faraday}) are
\begin{equation}
\frac{1}{r} \frac{\partial E_z}{\partial \theta}-\frac{\partial E_{\theta}}{\partial z} = -\frac{\partial B_r}{\partial t}
\end{equation}
\begin{equation}
\frac{\partial E_r}{\partial z}-\frac{\partial E_z}{\partial r}=-\frac{\partial B_{\theta}}{\partial t}
\end{equation}
\begin{equation}
\frac{1}{r} \left [\frac{\partial}{\partial r} \left (r E_{\theta}\right )-\frac{\partial E_r}{\partial \theta}\right ]=-\frac{\partial B_z}{\partial t}.
\end{equation}
For axisymmetric problems, the equations above can be  simplified. In particular, the current flows along the $\theta$ direction only and the magnetic field has the $r$ and $z$ components, therefore $B_{\theta}=0$; there is only one component ($E_{\theta}$) of the electric field (which is parallel to the current density $J_{\theta}$) and $E_r=E_z=0$; finally, all the derivatives with respect to $\theta$ can be put equal to zero.

Using ${\bf B} = \mu {\bf H}$, one is left with the following two governing equations
\begin{align}
-\frac{\partial E_{\theta}}{\partial z}&=\frac{\partial (\mu H_r)}{\partial t} \\
\frac{E_\theta}{r}+\frac{\partial E_{\theta}}{\partial r}&=-\frac{\partial (\mu H_r)}{\partial t}.
\end{align}
The current density is given by
\begin{equation}
J_{\theta}=
-\frac{\partial H_z}{\partial r}+\frac{\partial H_r}{\partial z}.
\end{equation}
\subsection{Extension to 3D}\label{sec:3D}
Certain applications cannot be accurately modeled in 2D and require the use of a full 3D model. Implementing equation~\ref{eq:faraday} in 3D is quite straightforward~\cite{Zhang:SST12}. 
One difference with respect to the 2D model is that the current density and the electric field do not necessarily need to be parallel, which allows exploring the so-called force-free configuration~\cite{Clem:PRB82}.
3D modeling can be used to explore the end effects in finite geometries, such as superconducting bulks~\cite{Ainslie:SST15} and stacks of HTS tapes~\cite{Zou:TAS16a}.

When cables are modeled in 3D, measures need to be taken in order to keep the size of the problem manageable. This typically involves avoiding the simulation of the whole geometry and focusing on a representative periodic `cell' of the cable. In that case, periodicity conditions need to be applied at the two ends of the simulated cell.

The 3D model also allows introducing spatial variation of the physical properties along the conductor's length. In particular, the effect of localized defects can be studied~\cite{Grilli:Cryo13}.

\subsection{Homogenized and multi-scale models}\label{sec:homogenization_multi_scale}
In certain applications, like large coils, the number of turns to simulate can be very large, in the range of hundreds or even thousands. Simulating all of them as individual objects can rapidly become a daunting task. One solution to the problem is to `homogenize' the cross section of the coil as one bulk conductor~\cite{Zermeno:JAP13}.
The superconductor's critical current density $J_{\rm c}$ is `diluted' to take into account the distance between the layers of superconducting material in the actual stack. In the case of HTS coated conductors, this distance includes all the composing layers (superconducting layers, substrate, buffer layers, stabilizer) as well as the separation between the tapes (e.g. electrical insulation). In the case of a coil, each tape in the simulated cross section represents a turn of the coil, and all the simulated tapes must carry the same current. This means that in the homogenized bulk the current cannot be let free to distribute in the whole cross section, but must be constrained to mirror the real situation, where each tape (turn) carries the same current.
In the numerical model, one must impose that the integral of the current density in the direction of the tape's width is always the same: this can be done by using an appropriate current constraint, as explained in~\cite{Zermeno:JAP13}.

The advantage of the homogenization becomes  evident for the simulation of large systems or when the distance between the turns becomes very small (so that it would require a very fine mesh if all the turns had to be simulated individually). For example, in~\cite{Queval:SST16} it was shown that, for the case of five 100-turn pancake coils, the homogenized model is 50 times faster than the model simulating all the tapes, with no appreciable difference (less than \SI{1}{\percent}) on the AC losses.

The homogenized problem is much simpler to solve because one can use a relatively coarse mesh (resulting in a much lower number of degrees of freedom) and less current constraints (whose number influences the speed of the solving process). The main drawback is that this method cannot take into account magnetization currents caused by a magnetic field parallel to the flat face of the tapes.

The homogenization method was extended to 3D in order to simulate a racetrack coil made of HTS coated conductors in~\cite{Zermeno:SST14a}. The idea is the same as in the 2D version of the method. However, the imposition of the constraint for having the same current flowing in each turn is more challenging. The current density has to follow the curvature of the tapes at the coils' ends. In~\cite{Zermeno:SST14a}, three strategies were proposed: 2D integral constraints, anisotropic resistivity, manual discretization of the anisotropic bulk. The last one was used because of better computational stability.

Another approach for modeling large numbers of interacting tapes is the multi-scale method~\cite{Queval:TAS13}: the idea is to simulate one conductor at a time using a boundary condition for the field it is subjected to as the result of the environment (for example, the field created by the other conductors). This magnetic field can be calculated in a relatively simple way, with magnetostatic models. If one is for example interested in calculating the AC losses of the whole device, the position of the tape of interest can then be moved and a `map' of the AC loss distribution in the device created. The advantage is that the simulations involving one tape are very fast and they can be truly parallelized. In addition, it is not necessary to consider all the positions of the tape in the device, but only some key ones. In~\cite{Queval:SST16},  it was shown that 25 equally distributed positions are used to produce a sufficiently accurate AC loss map of five 100-turn pancake coils. In the case considered in~\cite{Queval:SST16}, it was found that the current distribution used for the magnetostatic calculation has an important influence on the final AC loss value and hence on the accuracy of the results: for example, starting with a uniform current distribution in the superconductors is too rough an assumption and produces errors in the losses as large as~\SI{20}{\percent}.

\section{Applications}\label{sec:applications}

This section reviews the works that use the different forms of the $H$ formulation discussed in section~\ref{sec:Hform} for calculating AC losses in various applications and scenarios.

\subsection{Tapes and coils}
A challenge of modeling HTS coated conductor is the very large width-to-thickness ratio of the superconducting layer, which typically results in a very dense mesh and a very large number of degrees of freedom. In~\cite{Hong:JSNM10} it was found that artificially expanding the thickness (and correspondingly reducing $J_{\rm c}$) results in much faster simulation without compromising the accuracy of the results. In \cite{RodriguezZermeno:TAS11}, the authors proposed the use of a structured mesh for the superconducting layer, with elongated rectangular elements. This also allowed significantly reducing the size of the problem and is now the standard method for simulating coated conductors. A discretization of 50-100 elements along the tape's width is generally sufficient. Adding a discretization of a few elements along the superconducting layer's thickness does not usually result in an excessive number of degrees of freedom. Since in most cases a 1D description of the superconducting material is sufficient, one element along the thickness is often used.

Shen et al.~\cite{Shen:PhysC17} used the $H$ formulation  to calculate the eddy current AC loss in commercial non-stabilized and copper-stabilized HTS tapes, and they compared the results with experiments. They found that the eddy current losses in the copper have the expected dependence on frequency (power loss proportional to the square of the frequency) and give a substantial contribution from frequencies above~\SI{1}{\kilo\hertz}.

\begin{figure}[t!]
	\centering
	\includegraphics[width=8 cm] {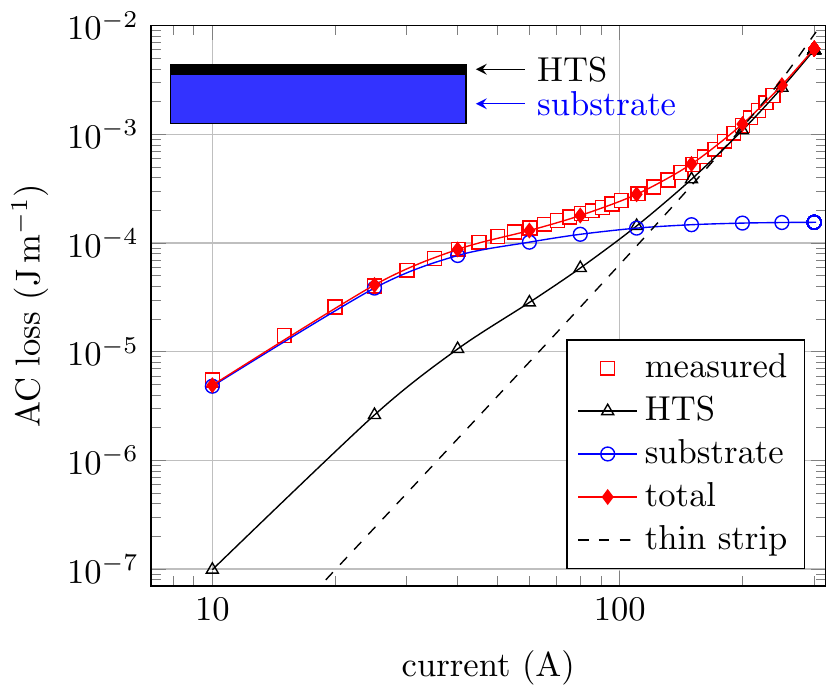}
	\caption{\label{fig:doan_new_with_strip}Transport losses in an HTS tape with magnetic substrate calculated by H formulation, showing the good agreement with measurement, and the separation of the losses into different components which are not directly available from measurements. Note that the magnetic substrate modifies the losses of the HTS material, which are different from those of a tape without magnetic substrate (indicated by the `thin strip' curve). Replotted from~\cite{Nguyen:SST10}.}
	\end{figure}
Nguyen et al.~\cite{Nguyen:SST10} calculated and measured the transport AC losses of HTS tapes with magnetic substrate, showing that at low current the ferromagnetic losses in the substrate give the more important contribution to the total losses (figure~\ref{fig:doan_new_with_strip}). As the current is increased, the ferromagnetic losses saturate and the losses of the superconducting layer are dominant.

 In~\cite{Nguyen:SST09,Nguyen:SST11} the $H$ formulation was used to calculate the AC losses of bifilar coils made of HTS tapes with and without magnetic substrate to be used for fault current limiter applications. The simulated configuration, consisting of a coil where the current of adjacent turns flows in opposite directions, is a typical example where the use of current constraints (see equation~(\ref{eq:current_constraints})) is particularly handy.
 
In~\cite{Zhang:SST12b}, Zhang et al. used an axisymmetric $H$ formulation model to calculate the AC losses in HTS pancake coils made of tape with magnetic substrate and compared the results with experimental measurements, finding good agreement. They also found that  the presence of a magnetic substrate does not influence the critical current of the coil; however, it affects the loss profile of each turn inside the coil under DC conditions, when the current approaches the critical current. This is due to the changes of the magnetic field profile caused by the magnetic substrate.

In~\cite{Ainslie:TAS13,Ainslie:TAS15}, Ainslie et al. used the 2D $H$ formulation to investigate the possibility of reducing the transport AC loss of HTS coils by means of flux diverters made with both weakly and strongly magnetic materials. The results show that significant loss reductions can be achieved and that the ideal diverter material should have a high saturation field and a low remnant field.

On the same topic, Liu et al~\cite{Liu:SST17} used numerical simulations to find the effects of the geometrical parameters of the flux diverters on the transport AC losses of an HTS coil.

In~\cite{deBruyn:TMAG16}, de Bruyn et al. presented a simplified method for calculating AC losses in stacks of superconducting tapes. The method aggregates the superconducting, substrate, and copper layers of several windings of the superconducting tape to form bulk elements. The resulting model is able to evaluate the AC losses in the superconducting layer faster than the full model. The accuracy of the proposed method depends on the current level at which the tapes operate.

Liang et al.~\cite{Liang:SST16} studied the AC losses in two types of low-inductance solenoidal HTS coils using 2D axisymmetric $H$ formulation and experiments, and proved that the braid type coil could be a suitable choice for superconducting fault current limiters as it has lower AC loss than the regular type coils.

Shen et al.\cite{Shen:TAS19b} used the 2D axisymmetric $H$ formulation to investigate the power dissipation circular HTS coils under the action of different oscillating fields and currents, and found that more AC losses were generated with the faster variation in waveform gradient within a certain time interval.

Zhang et al.~\cite{Zhang:SST12}  developed a 3D model of the $H$ formulation and used it for calculating the transport AC losses  of a cable composed of one layer of helically wound HTS tapes. They found that the transport losses increase with the twist angle, which they ascribed to the generation of an extra shielding current (caused by the twisting) flowing in the direction of transport current.

A full 3D FEM model of three twisted superconducting wires operating with AC transport current was presented in~\cite{Grilli:Cryo13}: the results show the effect of the twisting on the current density distribution and  agree well with those obtained with a 2D model that uses a change of coordinates to take the twist into account. In the same article, the potential of 3D simulations is shown for different examples:
i) the coupling currents between two thin superconducting filaments as a function of the resistivity of the material between them;
ii) the effect of a localized defect on the current flow in a two-filament structure;
iii) the effect of a varying cross section on the current distribution of a round wire.

Lyly et al.~\cite{Lyly:TAS13} used the 3D H formulation to model the losses of twisted NbTi wires with various wire geometries under different magnetic fields, and studied the operation conditions and coupling effects of filament bundles.

Lahtinen et al.~\cite{Lahtinen:SST12} compared three FEM methods, $A-V-J$ formulation, $T-\varphi$ formulation and $H$ formulation, for the calculation of hysteresis loss of a round superconducting wire, and the results showed that the $H$ formulation has the advantage of reasonable computation speed and boundary condition settings.

Lyly et al.~\cite{Lyly:TAS14} introduced a time harmonic method to calculate the eddy current losses in  twisted superconducting wires. The results were compared to those from the 3D $H$ formulation models, and good agreement was found in certain conditions.

\begin{figure}[t!]
	\centering
	\includegraphics[width=8 cm] {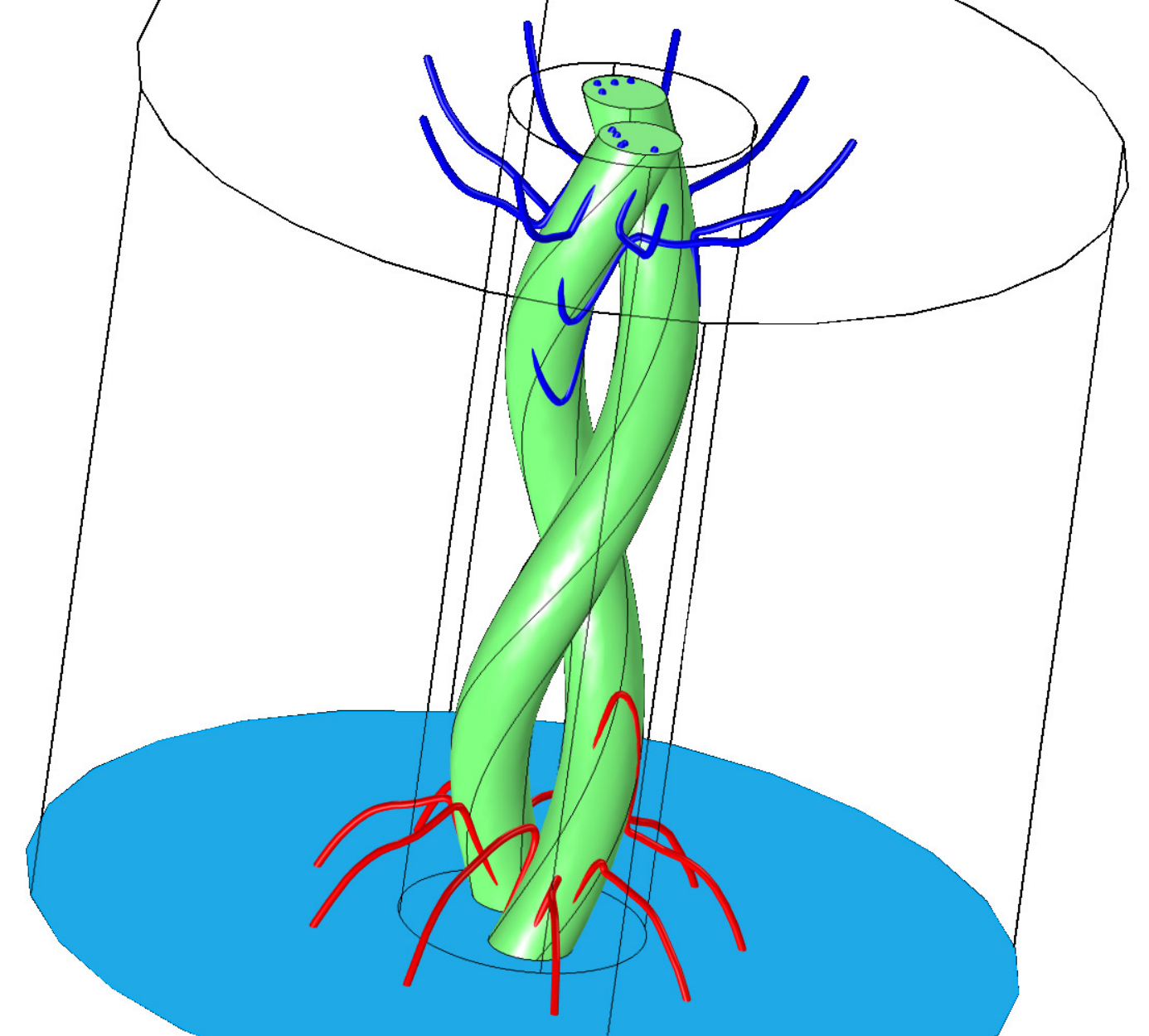}
	\caption{\label{fig:cohomology}3D simulations of two twisted superconducting filaments carrying AC transport current. The current streamlines demonstrate how part of the current can flow between the air to the superconductors. This can be avoided by increasing the resistivity of the air (which however tends to result in slower calculations) or by using cohomology functions in the so-called $H-\varphi-\varPsi$ formulation developed in~\cite{Stenvall:SST14}.}
	\end{figure}

Stenvall et al.~\cite{Stenvall:SST14} performed a similar comparison between the 3D $H$ formulation FEM model and a home-brewed code using  Gmsh with C{}\verb!++!, for the loss analysis of HTS twisted wire with varying applied magnetic field. They found that the $H$ formulation may underestimate the AC losses in the 3D HTS twisted wire model because of a current leakage to the air domain, if too low values of the air resistivity are used. An example is shown in figure~\ref{fig:cohomology}.

In~\cite{Zermeno:SST14a}, the 3D homogenization technique was used to model stacks and coils of HTS tapes. For validation, the 3D homogenization method was successfully tested against a 2D model considering all the individual conductors and enforcing all the individual currents. Both methods provided a remarkable good agreement over a large range of applied currents and over more than two orders of magnitude for the calculated AC losses. Then, the 3D model was used to simulate the more complex case of a racetrack coil, and it was found that the AC loss results agree with those of a planar 2D model only for low current fractions of $I_{\rm c}$. For
medium and high current, the losses calculated with the planar 2D model diverge and the use of the 3D homogenized model becomes necessary.

Eddy current models such as the $H$ formulation can in principle be used to estimate the losses caused by AC ripples on top of DC excitations in HTS. However, this kind of modeling presents some challenges, which have been described and discussed with examples in~\cite{Lahtinen:TAS13}.

Firstly, a finite $n$ exponent, even though as high as 40 (or, in general, a smooth $E-J$ constitutive law for the superconductor) causes the current penetration profile to 'relax' with time (going toward a homogeneous distribution), when there are no longer any changes in the net current or external field. This leads to very slowly descending loss curves and the evaluation of the cyclic losses caused by the ripples depends on where along those curves the evaluation is performed. This was confirmed by  simulations of a coated conductor carrying DC current with AC ripples in~\cite{Xu:SST15}, where the details of the current density distribution along the tape's width at different instants were studied and put in relation to the instantaneous power dissipation.

Secondly, in the case of a multi-filamentary wire modeled in 2D, the requirement of filaments being uncoupled with respect to the external field is contradictory with the requirement of parallel connected filaments in the eddy current models, although this obstacle could be surpassed by using current constraints.

In~\cite{Hong:TAS11b}, the authors investigated the effects of AC ripple fields on an HTS racetrack coil subjected to DC background field or DC transport current, two conditions that can be met in the field windings of electrical machines. The results show that both the DC background field and transport current would significantly increase the values of AC losses, especially when the AC ripple field is small.
Similar results were found in~\cite{Ying:TAS13}, where different ratios of DC and AC currents were simulated and measured. The authors also found that a small value of DC offset in AC current would not affect the AC losses.

In~\cite{Lahtinen:JAP14}, the authors presented a comprehensive study of the AC ripple losses of an HTS coated conductor for a variety of combinations of AC and DC background fields and transport current. They used two numerical models -- the $H$ formulation with power-law and the Minimum Magnetic Energy Variation (MMEV) method with the critical state -- and performed experimental measurements. For pure AC cases and DC-AC cases with the AC field significant enough compared to the DC field, the agreement between the models (and with measurements) is good.
For some of the studied cases, CSM and ECM yield different predictions for the behavior of DC biased superconductors, which are ascribed to the different $E-J$ relations used in the models -- see also~\cite{Lahtinen:TAS13} discussed above.
The experiments seem to suggest that, for a given current density $J$, the electric field $E$ is lower than that predicted with the power-law $E-J$ relation, and therefore indicate an $E-J$ relation closer to the critical state model for low E. However, the authors acknowledge that further research on the topic -- e.g. by means of higher resolution field map measurements --  is needed, because neither the power-law nor the critical state model are able to predict the observed behavior.

\subsection{High-current cables}
The $H$ formulation has been used to calculate the AC losses in different types of cables, composed of HTS tapes.
\subsubsection{Roebel cables}
Roebel cables are made of intertwined meander-shaped strands obtained from HTS coated conductors~\cite{Goldacker:SST14}. For most AC loss calculation purposes, it is sufficient to consider their transversal cross section, so that the cables can be simulated as two stacks of tapes~\cite{Terzieva:SST10}.

In~\cite{Grilli:SST10b}, the authors used this 2D approximation to calculate the transport and magnetization losses. The results of the $H$ formulation were successfully compared with those obtained with the MMEV model. Two different scenarios of strand coupling were considered, which provided two limits for the AC losses of the actual cable.
In~\cite{Pardo:SST12a}, the same authors extended their models to take into account the angular dependence of the critical current density on the magnetic field. Situations where the anisotropy of such dependence can play a role were identified.

The AC losses of a Roebel cable under the simultaneous action of AC transport current and AC magnetic field were calculated in~\cite{Grilli:TAS16b}. Two cables made of tapes from different manufacturers, characterized by different angular dependence of $J_{\rm c}$, were analyzed and the AC loss results compared with experimental data obtained with a calorimetric method measuring the evaporation of liquid nitrogen.

In~\cite{Thakur:SST11b}, the authors studied the frequency dependence of the transport losses of an 8-strand Roebel cable by simulations and experiments. The frequency was varied between \SI{50}{\hertz} and \SI{10}{\kilo\hertz}. They found that for low and medium current amplitudes the AC losses per cycle decrease as $f^{-2/n}$ with $n=26$, whereas for high current amplitudes they are proportional to $1/f$.

In~\cite{Grilli:TAS13,Grilli:TAS14b}, the authors studied the transport losses of different pancake coils assembled from the same \SI{5}{\meter}-long Roebel cable. The coils differ in terms of number of turns and turn-to-turn separation. The experiments revealed that at low and medium current amplitudes the losses are dominated by the dissipation occurring in the copper current lead used to inject the current in the cable. Once this contribution was taken into account by means of a dedicated 3D simulation and added to the results obtained for the coils with the $H$ formulation, it was possible to match the experimental data.

In~\cite{Zermeno:SST13}, Zermeno et al. developed a full 3D H formulation FEM model of a Roebel cable with 14 strands. The 3D model simulates a periodic cell representing the various positions that a given strand occupies along the length of the cable. The calculated AC losses were similar to those obtained with 2D simulations. The 3D model revealed the presence of high dissipation near the corners of the strands. Although this localized dissipation does not contribute significantly to the cyclic losses of the whole cable, it can represent a potential stability issue for this type of cable. To this date, this is the only full 3D model of a Roebel cable, where also the thickness of the superconductor layer is simulated.
\subsubsection{CORC\textsuperscript{\textregistered} cables}
CORC\textsuperscript{\textregistered} cables are assembled by winding multiple layers of HTS coated conductor on a narrow former~\cite{VanderLaan:SST19}. 

In~\cite{Majoros:SST14}, Majoros et al. investigated the magnetization AC losses and heat generation in the HTS CORC\textsuperscript{\textregistered} cables by experiments and 2D $H$ formulation FEM models. In the simulations they observed the screening effects that make the AC losses decrease with higher frequencies. 

Sheng et al.~\cite{Sheng:TAS17} developed a 3D $H$ formulation model for computing the magnetization AC losses in a CORC\textsuperscript{\textregistered} cable, which can account for fully coupled or fully uncoupled strands. The latter situation provided a better match with experimental results. The simulations also suggested that a stronger shielding effect can be achieved by increasing the coverage ratio of the HTS tapes and that the end effects  cannot be omitted in the study of magnetization loss of short CORC\textsuperscript{\textregistered} cable. In particular, a correction coefficient should be used when  experimental results of short CORC\textsuperscript{\textregistered} samples are used to predict the magnetization loss of very long cables.

Terzioglu et al.~\cite{Terzioglu:SST17} studied experimentally the AC losses  of a CORC\textsuperscript{\textregistered} cable with copper former caused by transport AC current, external AC field and their combination. 
They found that the magnetization AC losses increase due to losses in the metallic former, but that, at low field amplitudes, the magnetization AC loss of the complete cable is lower than the loss in the bare former. They ascribed this result to the shielding of the magnetic field by the superconductor, and they used numerical simulations based on the 3D $H$ formulation to support this explanation. The authors also suggest that the losses can be reduced by using a material with low electrical conductivity and high thermal conductivity for the former.

\subsubsection{Stacked-tape cables}

In~\cite{Ainslie:PhysC12}, Ainslie et al. studied the transport AC loss of stacks of HTS coated conductors, with and without magnetic substrate.
When investigating the effect of a magnetic substrate, it was found that the transport AC loss is significantly increased, especially in the central region of the stack, due to an increased localized magnetic flux density. The ferromagnetic loss of the substrate itself is found to be negligible in most cases, except for small magnitudes of current where the substrate is
not yet saturated.

In~\cite{Krueger:TAS15, Grilli:PhysC15}, the authors showed by simulations that in the case of a twisted HTS stacked-tape cable subjected to a transversal AC field, the magnetization losses are determined by the field component perpendicular to the tape. As a consequence, they can be calculated on a straight HTS stacked-tape cable (which can be simulated in 2D) and scaled by a factor $2/\pi$. Termination resistances can be accounted for by simulating 
two non-connected domains~\cite{Zermeno:SST14b}, one for the  resistances and one for the HTS tapes. The current is imposed on one side of the  resistances (in parallel). On the other side, the current exiting a particular  resistance is made flow into the corresponding tape by means of appropriate constraints. The model for the termination resistances requires only a minimal geometrical layout with a small mesh in the discretization and a consequently low number of degrees of freedom. In the specific case of a single HTS stacked-tape, the 3D model is not necessary because the contribution of contact resistances can be directly inserted in a 2D model~\cite{Zermeno:SST14b}. However, this approach can be of interest for more complex situations for which the 2D approximation is not valid.

In~\cite{Kan:TAS18}, Kan et al. carried out an AC loss study of quasi-isotropic strands cable manufactured by HTS. The idea is to build a stacked-tape cable with tapes oriented in two different ways, in order to reduce the effects of the strongly anisotropic angular dependence of $J_{\rm c}$ typical of HTS coated conductors. They calculated the magnetization AC loss at \SI{4.2}{\kelvin} and \SI{77}{\kelvin}.

In~\cite{Shen:TAS19}, Shen et al. used 2D simulations to investigate the applicability of the cross-conductor concept to cables for AC power transmission. They simulated single-phase and three-phase cables and found out that, in order to operate with an acceptable level of AC losses,  these cables can only operate at a low fraction of their critical current.
\subsubsection{3D modeling of cables with twisted structures}
In~\cite{Makong:TMAG18b}, Makong et al. developed a simplified version of the 3D H formulation FEM model (implemented in GetDP) to simulate the performance and calculate the AC loss of twisted multi-filamentary HTS wires subjected to transverse magnetic fields. The adopted geometric transformation allows studying the wires in the Frenet frame moving along the helical trajectory resulting from twisting filaments and saves considerable computation time. The accuracy of the method showed some dependence on the twist pitch and on the magnitude of the external field.

In~\cite{Escamez:TAS16}, Escamez et al. used the 3D $H$ formulation implemented in Daryl Maxwell  to analyze the AC losses of the $\rm MgB_2$ wires with 6, 18 and 36 filaments. The authors demonstrated -- by means of comparison with 2D models -- the necessity of a 3D model in order to compute correctly the coupling losses, which contribute significantly to the losses, especially for high applied field values. They also considered the impact of the nonlinear permeability of the nickel matrix, which generates substantial additional losses. The authors underlined the necessity of speeding up the solving process for simulations of such complexity, especially for the 36-filament case.

\subsection{High field magnets and power applications}
\subsubsection{High-field magnets}
Xia et al.~\cite{Xia:SST15} used the homogenized $H$ formulation to investigate the electromagnetic behavior of the HTS prototype coils of the National High Magnetic Field Laboratory 32 T all-superconducting magnet. Beside AC loss calculation, that approach can be used to study the central magnetic field drift, magnetic field deviation, and to optimize the design of high field coils.

\begin{figure}[t!]
	\centering
		\includegraphics[width=12 cm] {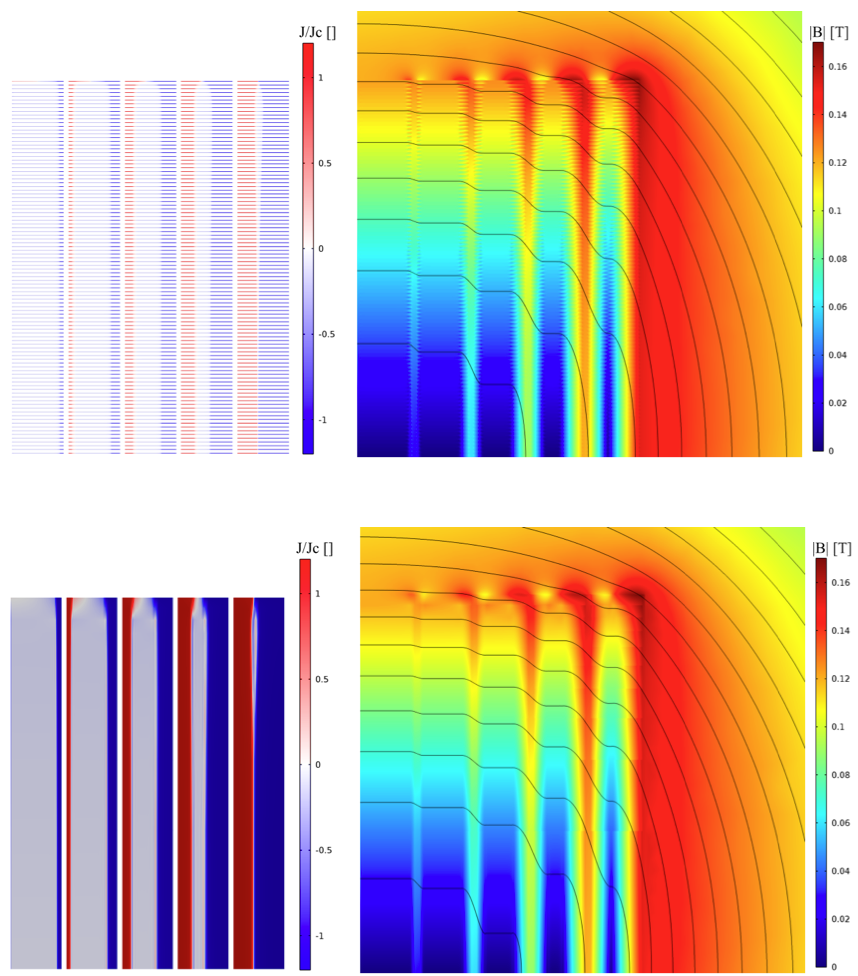}
	\caption{\label{fig:Loic}Simulation of a matrix of 10$\times$200 tapes, representative of the turns in the cross section of a high-field magnet. Due to the symmetry of the problem, only 5$\times$100 tapes are simulated. The top figures represent the current density and magnetic flux density distributions obtained by simulating all tapes with the time-dependent $H$ formulation of Maxwell's equations~\cite{Brambilla:SST07}. The bottom figures represent the same quantities obtained with a homogeneous bulk approximation: the results are very similar, but the computation is  50-60 times faster. Reprinted with permission from~\cite{Queval:SST16}.}	
\end{figure}

Qu\'eval et al.~\cite{Queval:SST16} used the $H$ formulation to calculate the losses of a large coil composed of ten 200-turn pancakes (reduced to five 100-turn pancakes thanks to symmetries). They used both the homogenized and multi-scale approaches and compared the results with the reference case obtained by simulating all the tapes. The homogenized and multi-scale approaches allow saving a great amount of time, with no significant loss of accuracy. In the considered case, the homogenization resulted to be the faster (50-60 times faster than the reference case), although the full potential of the multi-scale approach obtainable with full parallelization was not tested. An example is shown in figure~\ref{fig:Loic}.

\subsubsection{Electrical machines}
Ainslie et al.~\cite{Ainslie:SST11} used an improved $H$ formulation FEM model with structured mesh to calculate the transport AC loss in the HTS coils for large electric machines, including the loss contribution of the ferromagnetic tape substrate.

Zhang et al.~\cite{Zhang:TAS13} calculated the AC losses in HTS racetrack coils to be used as  armature windings in electrical machines. The 2D $H$ formulation model, which considers the tapes' anisotropic angular dependence of $J_{\rm c}$ on the magnetic field, was successfully validated  against experimental measurements of AC losses performed with both the  electric and calorimetric methods. The authors calculated the transport loss of the HTS armature winding in terms of electrical loading of the machine, and pointed out that a distributed winding would be a feasible way to reduce transport loss.

Zhang et al.~\cite{Zhang:SST15} performed a calorimetric measurement of  the total AC loss of a 2G HTS racetrack coil subjected to both an applied current and an external magnetic field. They found that when the external magnetic field is perpendicular to the tape surface, the total AC loss show a modulation in regard to the phase shift between the applied current and external magnetic field. The authors then used the $H$ formulation to support the experimental data and to study in detail  the influence of the phase shift between the current and the field on the total AC loss.

Qu\'eval et al.~\cite{Queval:TAS13} used the $H$ formulation FEM model to estimate the AC losses of a superconducting wind turbine generator (employing Bi-2223 tapes) connected to the grid.  For this study, the authors performed single-tape simulations on a fraction of the tapes of the coils' cross section. They calculated the magnetic field impinging on the tape by means of an unidirectional coupling between the machine model and the HTS tape model. With this approximation, they found that the steady-state AC loss of a \SI{10}{\mega\watt}-class wind turbine generator could be under~\SI{60}{\watt}. 

Li et al.~\cite{Li:TAS16b} did a study of HTS armature windings in a \SI{15}{\kilo\watt}-class fully HTS synchronous generator. They used the 2D $H$ formulation on a simplified geometry and, by applying the minimum and maximum homogenous leakage magnetic fields to the boundary of the simulated domain, they estimated the lower and upper limits for the AC losses.

In~\cite{deBruyn:SST17}, de Bruyn et al. used the homogenized $H$ formulation to predict AC losses resulting from non-sinusoidal transport currents as are present in highly dynamic motors with AC armature coils, finding good agreement with experimental data.

\subsubsection{Fault current limiters}
Although modeling the current limiting behavior of superconducting fault current limiters requires thermal models, electromagnetic calculations at constant temperature can be used to evaluate the AC losses of these devices during normal operation.

Hong et al.~\cite{Hong:TAS12} investigated the AC loss in a 10 kV resistive-type superconducting fault current limiter, using both experiments and $H$ formulation calculations. The numerical model was used to calculate the AC loss during normal operation, for different values of the transport current. Good agreement was found between measurements and simulations for currents above \SI{50}{\percent} of the critical current.

Jia et al.~\cite{Jia:TAS17} performed a numerical analysis of a saturated core type superconducting fault current limiter and partly with loss analysis using $H$ formulation. They found that in the DC biasing coil the outermost HTS layers have much higher losses than the central layers, and therefore the quench is more likely to occur in the outermost layers. In addition, the total loss can be estimated by interpolating the loss in several key layers in the outermost, middle, and innermost layers.

Shen et al.~\cite{Shen:TAS19} used the $H$ formulation  to investigate the details of the  power dissipation of a three-phase \SI{35}{\kilo\volt}/90~MVA saturated iron core superconducting fault current limiter. The estimated losses are up to the \SI{}{\kilo\watt} level and can be even higher, depending on the amplitude of the used DC bias current, which should therefore be taken into high consideration during the design of these devices.

\subsubsection{Transformers and SMES}

Song et al.~\cite{Song:Cryo18} used the $H$ formulation with homogenization to calculate the AC losses of a 1 MVA HTS transformer with approximately one thousand turns in the HV winding and with solenoid LV windings, each phase with 20 turns of 15/5 Roebel cable. They also modeled a standalone solenoid coil with the same geometry as the LV winding.
The authors successfully reproduced the main features of AC losses as well as the current and the magnetic flux distributions that were previously calculated with the MMEV method in~\cite{Pardo:SST15b}. The disagreement, at rated current, between the AC loss estimation with the $H$ formulation and the experimental results as well as the predictions of the MMEV method is less than 20\% without optimizing the mesh.

In~\cite{Wang:TAS17b}, Wang et al. did an AC loss study of a hybrid HTS magnet of approximately 7000 turns  for superconducting magnetic energy storage (SMES) by using the $H$ formulation with homogenization method, which was first validated by comparing its results with those of the full model for both YBCO coils and BSCCO coils of smaller size. The AC losses of the SMES were studied and analyzed during various power exchange conditions. The results show that AC loss is concentrated at the top and bottom ends of the magnet and that the higher the steady-state current, dynamic current, and current ramp rate, the higher the AC loss power. The authors also found that, for the SMES under study, a steady-state current of less than \SI{100}{\ampere} could be an appropriate choice for reducing AC loss.
 
In~\cite{Morandi:TAS16}, Morandi et al. used 2D simulations of a Roebel cable to estimate the dissipation occurring during the charging and discharging of a \SI{1}{\mega\watt}/\SI{5}{\second} SMES with solenoidal and toroidal geometry. The simulation showed that the losses are dominated by the applied field. As a figure of merit for comparing the loss performance of the two topologies, the authors combined the loss per unit length corresponding to each of the levels of perpendicular field with the length of conductor exposed to that field, thus obtaining a quantitative ``loss indicator'', which resulted to be higher for the toroidal geometry.

\section{Discussion}\label{sec:discussion}
As demonstrated in the previous sections, the $H$ formulation FEM model has shown its powerful capability to estimate the AC losses for a wide range of HTS topologies and in a large variety of operating scenarios. The overwhelming majority of results published in the literature uses the implementation in the FEM software package COMSOL Multiphysics. In truth, other implementations of the $H$ formulation FEM model have been proposed. These include commercial software packages, like FlexPDE~\cite{Lorin:TAS15} and Matlab~\cite{Lahtinen:SST12}, open-source environments like GetDP~\cite{gitlabweb}, and home-made FEM codes like Daryl Maxwell~\cite{Escamez:TAS16}. However, they represent a minority. It would be na\"{i}ve not to recognize that one of the reasons (if not the main one) of the popularity of the use of COMSOL Multiphysics is the easiness of implementation of the model. In recent years, things have been made even easier thanks to the built-in `MFH module', where equation~(\ref{eq:faraday}) is already implemented, and one does not need to write it in the `General PDE module' for partial differential equations anymore.

The easiness of implementation is particularly attractive because it allows new users to get up to speed in a relatively short time. In research groups, it also allows a rapid and efficient transfer of knowledge when personnel come and leave.
 
Another reason of the popularity is that it is quite easy to exchange models between users, and that the files for numerous topologies and application scenarios are publicly available~\cite{nummodweb}. These include not only the basic implementation of the model, but quite advanced models for cables and windings. 

In terms of accuracy when compared against experimental data, the $H$ formulation has a performance that is consistent with that of similar numerical models. In general, the calculated AC losses are in good agreement with the experiments, and the accuracy can be quantified as varying between \SI{10}{\percent} and \SI{50}{\percent}. This can be considered satisfactory, for several reasons:
\begin{itemize}
\item It is not always possible to have a direct precise characterization of the properties of the superconductor tape under analysis, for example the angular dependence of $J_{\rm c}$ on the magnetic field;
\item The tapes' properties are almost always assumed to be uniform along the tape's length, but variations in the order of 5-\SI{10}{\percent} are very common;
\item In the case of tape assemblies like cables and coils, the tapes present some degree of misalignment with respect to their theoretical position;
\item The AC loss measurements are prone to significant disturbances of different type, for example: spurious electromagnetic signals, difficulty in isolating a (very often small) voltage component in phase with a reference signal, extremely low levels of evaporated cryogenic liquid, etc.
\end{itemize}
The $H$ formulation has also important drawbacks. In terms of use, the implementation in COMSOL Multiphysics, has some of the typical problems related to using a commercial software. The code is not accessible, and for certain aspects the code remains a sort of `black box' to the user. In addition, the license price could be not affordable for everyone. On the other hand, the implementation in open-source codes is not trivial, and in general it is very time consuming, especially for first-time users.
In terms of computational efficiency, the $H$ formulation FEM model is not the best choice. This has primarily to do with the fact that the air domains need to be simulated as well, which may waste a lot of degrees of freedom.  For example, in~\cite{Grilli:SST10b} the computation times of the $H$ formulation were reported to be one order of magnitude longer than those obtained with a homemade code based on a variational method. On the other hand, in~\cite{Sirois:JPCS08} the $H$ formulation was reported to be faster than another FEM method based on $T-\Omega$  formulation implemented in FLUX.
Another pitfall is that the $H$ formulation FEM model implemented in COMSOL Multiphysics does not take advantage of computing parallelization. So, having access to large computer clusters does not necessarily help.
For the simulation of complex systems with coated conductors, the recently developed $T-A$ formulation seems to be a better alternative~\cite{Zhang:SST17}, as long as the superconductor layer can be approximated as an infinitely thin object.
\section{Conclusion}\label{sec:conclusion}
This article reviewed the calculation of the AC losses for various HTS topologies based on the $H$ formulation FEM model. A massive number of studies demonstrate that the $H$ formulation is one of the most widespread models used to calculate AC losses in HTS and has become the de facto standard numerical tool for that purpose. Among the reasons of this success is the easiness of implementation in commercial software, although the formulation has also been successfully implemented in home-made codes.
The early models for individual conductors have been improved and extended to be able to simulate scenarios of increasing complexity: from applications with nonlinear magnetic materials, to cables, medium-size coils, and large superconducting magnet systems. For large superconducting magnet systems, the models have been adapted to avoid the simulation of all the individual tapes, by means of techniques such as the multi-scale and the homogenization methods.
For the simulation of thin superconductors, like HTS coated conductor, the $H$ formulation is now facing some serious competition from the recently developed $T-A$ formulation. However, given its flexibility of application to different scenarios (including some that cannot be handled by the $T-A$ formulation), it is likely that the $H$ formulation will retain the leading role among the simulation tools for AC loss calculation in HTS in the years to come.

\providecommand{\newblock}{}

\end{document}